\def\aa{{A\&A}}

\def\aj{{AJ}}
\def\annrev{{ARA\&A}}
\def\apj{{ApJ}}
\def\apjs{{ApJS}}

\def\mnras{{MNRAS}}

\def\prd{{Phys. Rev. D}}

\documentclass[cup5b]{caps}
\usepackage{graphicx}
\usepackage{amssymb}
\usepackage{ociwsymp1}
\HeadText{S. L. Shapiro}

\begin{document}

\pagenumbering{arabic}

\author[]{S. L. SHAPIRO\\The University of Illinois at 
Urbana--Champaign}

\chapter{Formation of Supermassive Black Holes: Simulations in
General Relativity}

\begin{abstract}
There is compelling evidence that supermassive black holes exist.  Yet the 
origin of these objects, or their seeds, is still unknown.  We discuss several 
plausible scenarios for forming the seeds of supermassive black holes. These 
include the catastrophic collapse of supermassive stars, the collapse of 
relativistic clusters of collisionless particles or stars, the gravothermal 
evolution of dense clusters of ordinary stars or stellar-mass compact objects, 
and the gravothermal evolution of self-interacting dark matter halos. 
Einstein's equations of general relativity are required to describe key 
facets of these scenarios, and large-scale numerical simulations are performed 
to solve them.
\end{abstract}

\section{Introduction}

There is substantial evidence that supermassive black holes (SMBHs) of mass 
$\sim 10^6 - 10^{10} ~M_{\odot}$ exist and are the engines that power active 
galactic nuclei (AGNs) and quasars (Rees 1998, 2001; Macchetto 1999).  There 
is also ample evidence that SMBHs reside at the centers of many, and perhaps 
most, galaxies (Richstone et al. 1998; Ho 1999), including the Milky Way 
(Genzel et al. 1997; Ghez et al. 2000; Sch\"odel et al. 2002). 

\begin{figure}[tb]
\centering
\includegraphics[height=7cm]{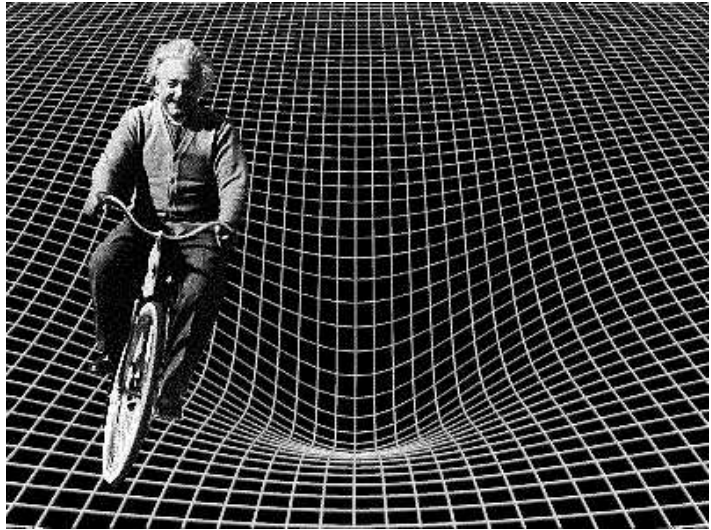}
\caption{The formation of a black hole is a strong-field gravitational 
phenomenon in curved spacetime that requires Einstein's equations of general 
relativity for a description and, in nontrivial cases, numerical simulations 
for a solution.}
\label{eins}
\end{figure}

Since quasars have been discovered out to redshift $z \gtrsim 6$ (Fan et al. 
2000, 2001), the first SMBHs must have formed by $z_{\rm BH} \gtrsim 6$, or 
within $t_{\rm BH} \lesssim 10^9$ yrs after the Big Bang.  However, the 
cosmological origin of SMBHs is not known. This issue remains one of the 
crucial, unresolved components of structure formation in the early universe.
Gravitationally, black holes are strong-field objects whose properties are 
governed by Einstein's theory of relativistic gravitation --- general 
relativity.  General relativistic simulations of gravitational collapse to 
black holes therefore may help reveal how, when and where SMBHs, or their 
seeds, form in the universe. Simulating plausible paths by which the first 
seed black holes may have arisen is the underlying motivation of our 
investigation (see Fig.~\ref{eins}).

\section{The Boltzmann Equation}
 
Various routes have been proposed over the years by which SMBHs or their seeds 
might arise by conventional physical processes (see, e.g., Fig. 1 in Rees 
1984).  Some routes are hydrodynamical in nature, such as the formation and 
collapse of supermassive stars (SMSs), while others are stellar dynamical, 
like the evolution and collapse of collisionless clusters. The Boltzmann 
equation provides a common mathematical framework for comparing competing 
scenarios:
\begin{equation} \label{boltz}
\frac {Df}{Dt} = \left(\frac{\partial f}{\partial t}\right)_{\rm collisions}.
\end{equation}
In equation~(\ref{boltz}) $f$ is the phase-space distribution function for
the matter, which might be in the form of a gaseous fluid, collisionless
particles, and/or stars. The left-hand side of the equation represents the
total time derivative of $f$ following a matter element along its trajectory 
in phase space. The right-hand side describes the role of collisions in 
modifying the phase-space distribution along the trajectory. To treat different
scenarios the Boltzmann equation must be solved in different physical regimes, 
all of which share gravitation as the the dominant long-range interaction. 

Table 1.1 summarizes some of the SMBH formation simulations that we have
performed in recent years. Typically, every scenario falls into one of three 
distinct regimes. In the Vlasov (collisionless Boltzmann) regime the dynamical 
timescale of the system, $t_d$, which is the time for matter to cross from one 
side of the system to the other, as well as the time to achieve virial 
equilibrium by violent relaxation, is much shorter than the relaxation 
timescale $t_r$, the time for  the system to reach thermal equilibrium via 
collisions. In pure Vlasov simulations the integration time $t$ may exceed 
$t_d$ but always remains much shorter than $t_r$. In such cases collisions can 
be ignored. The system can be evolved to dynamical (virial) equilibrium, but 
not thermal equilibrium. In the secularly collisional regime, $t_d$ again is 
much shorter than $t_r$ but the integration time is now much longer than 
$t_r$. Here collisions are crucial in driving the quasi-stationary evolution 
of the virialized system.  Since the timescale for collisions remains much 
longer than the dynamical timescale, collisions can often be handled  
perturbatively by tracking the secular drift of the system from one, nearly 
collisionless, virialized state to the next.  This is the approach adopted in 
the Fokker-Planck approximation to the Boltzmann equation. In the 
collision-dominated regime, $t_r$ is much shorter than $t_d$ and the system 
behaves as a fluid. This regime embraces all of hydrodynamics. 

\begin{flushleft}
\begin{tabular}{c}
Table 1.1 Boltzmann simulations of SMBH formation scenarios\\
{\sf\scriptsize
\begin{tabular}{lccc} \hline
\ \ REGIME & 
\begin{tabular}{c}
Vlasov\\ 
(Collisionless)
\end{tabular} &
\begin{tabular}{c}Fokker-Planck\\ (Secularly Collisional)\end{tabular} &
\begin{tabular}{c}Fluid\\(Collision Dominated)\end{tabular}\\\hline\hline
\begin{tabular}{l} TIME SCALE\\ ORDERING\end{tabular}& $t_r >> t >> t_d$ &
$t >> t_r >> t_d$ & $t >> t_d >> t_r$\\ \hline
\ \ SCENARIOS & \begin{tabular}{c} dynamical\\collapse of\\a
relativistic\\cluster of\\(1)\ compact stars\\ (NSs or \\stellar-mass
BHs)\\or\\(2)\ collisionless$\quad$\\particles\end{tabular}&\begin{tabular}{c}
``gravothermal\\catastrophe''\\drives core\\contraction of\\ (1)\ a dense
cluster\\of compact stars;\\or\\(2)\ a dense cluster\\of ordinary
stars;\\or\\(3)\ an SIDM halo\end{tabular} & \begin{tabular}{c}
hydrodynamical\\collapse\\of an SMS\end{tabular}\\ \hline
\ \ GRAVITATION & GR & Newtonian & GR, PN$^*$ \\ \hline
\begin{tabular}{l} SPATIAL \\ SYMMETRY \end{tabular} & \begin{tabular}{c}
Spherical;\\ Axisymmetrical \end{tabular} & Spherical & \begin{tabular}{c}
Spherical;\\ Axisymmetrical;\\Arbitrary$^*$\end{tabular} \\ \hline
\begin{tabular}{l} COMPUTATIONAL\\ DIMENSIONS \end{tabular} &
\begin{tabular}{c} 1 + 1;\\ 2 + 1 \end{tabular} & 1 + 1 &
\begin{tabular}{c}
1 + 1; \\ 2 + 1;\\ 3 + 1$^*$\end{tabular} \\ \hline
\begin{tabular}{l} COMPUTATIONAL \\ TECHNIQUE \end{tabular} &
\begin{tabular}{c} particle simulation \\ (matter)\\ + \\
finite-differencing \\ (field)\end{tabular} & finite-differencing &
finite-differencing \\ \hline
\end{tabular}}
\end{tabular}
\end{flushleft}

Scenarios considered to date for forming SMBHs, or their seeds, in the Vlasov 
regime focus on the dynamical collapse of a radially unstable, relativistic 
cluster of compact stars (neutrons stars or stellar-mass black holes) or 
collisionless particles.  Scenarios in the secularly collisional regime 
typically involve the gravothermal contraction of a dense cluster of ordinary  
stars or compact stars which undergo collisions and mergers, leading to a 
build-up of massive black holes.  The gravothermal contraction of a 
self-interacting dark matter halo (SIDM) in the early universe may also 
produce a SMBH.  Hydrodynamical scenarios typically focus on the collapse of a 
SMS or gas cloud.  Not surprisingly, simulations performed in the different 
regimes require very different computational approaches.  Table 1 indicates
that the different scenarios have been tackled by adopting various degrees of 
spatial symmetry to simplify the calculations. A summary of the results
of these simulations is given in the sections below.

\section{Numerical Relativity}

Numerical relativity --- the art and science of developing computer algorithms 
to solve Einstein's field equations of general relativity --- is the principal 
tool needed to simulate plausible black hole formation processes.  The 
underlying equations are multidimensional, highly nonlinear, coupled partial 
differential equations in space and time. They have in common with other areas 
of computational physics, like fluid dynamics and MHD, all of the usual 
problems associated with solving such nontrivial equations. However, solving
Einstein's equations poses some additional complications that are unique to
general relativity. The first complication concerns the choice of coordinates. 
In general relativity, coordinates are merely labels that distinguish points
in spacetime; by themselves coordinate intervals have no physical significance. 
To use coordinate intervals to determine physically measurable (proper) 
distances and times requires the spacetime metric, but the metric is 
determined only after Einstein's equations have been solved. Moreover, as the 
integrations proceed, it often turns out that the original (arbitrary) choice 
of coordinates turns out to be bad, because, for example, singularities 
eventually are encountered in the equations.  The gauge freedom inherent in 
general relativity --- the ability to choose coordinates in an arbitrary way 
--- is not always easy to exploit successfully in a numerical routine.  

The appearance of black holes always poses a complication in a numerical 
relativity simulation.  Black holes inevitably contain spacetime singularities 
--- regions where the gravitational tidal field, the matter density, and the 
spacetime curvature all become infinite. Encountering such singularities 
results in some of the terms in Einstein's equations becoming
infinite, causing overflows in the computer output and premature termination
of the numerical integration. Thus, when dealing with black holes, it is
crucial to choose a technique which avoids the spacetime singularities inside.
Some of the techniques involve choosing appropriate coordinate gauges that
avoid or postpone the appearance of singularities inside black holes. Others
involve excising the black hole interiors altogether from the numerical grid.

One of the main goals of a numerical relativity simulation is to determine the 
gravitational radiation generated from a dynamical scenario.  However, the 
gravitational wave components usually constitute small fractions of the 
background spacetime metric. Moreover, to extract the waves from the
background requires that one probe the spacetime in the far-field or radiation
zone, which is typically at large distance from the strong-field central 
source.  Yet it is the strong-field, near-zone region that usually consumes 
most the computational resources (e.g., spatial grid) to guarantee accuracy. 
Furthermore, waiting for the wave to propagate to the far-field region usually 
takes nonnegligible integration time. Overcoming these difficulties to 
reliably measure the wave content thus requires that a code successfully cope 
with the problem of dynamic range inherent in such a simulation.

For a recent review of the status of numerical relativity, and a summary of 
the key equations, see Baumgarte \& Shapiro (2003) and references therein.

\section{Collapse of a Rotating SMS to a SMBH}

SMBHs must be present by $z_{\rm BH} \gtrsim 6$ to power quasars.  It has been 
suggested (Gnedin 2001) that even if they grew by accretion from smaller 
seeds, SMBH seeds $\gtrsim 10^5 ~M_{\odot}$ must have formed at $z \approx 9$ 
to have had sufficient time to build up to a typical mass of 
$\sim 10^9 ~M_{\odot}$. A likely progenitor is a very massive object (e.g., an 
SMS) supported by radiation pressure. 

SMSs ($10^3 \lesssim M/M_{\odot} \lesssim 10^{13}$) may form when contracting 
or colliding primordial gas builds up sufficient radiation pressure to inhibit 
fragmentation and prevent star formation (see, e.g., Bromm \& Loeb 2003).  
SMSs supported by radiation pressure will evolve in a quasi-stationary
manner to the point of onset of dynamical collapse due to general relativity
(Chandrasekhar 1964a,b; Feynmann, unpublished, as quoted in Fowler 1964).
Unstable SMSs with $M \gtrsim 10^5 ~M_{\odot}$ and metallicity 
$Z \lesssim 0.005$ do not disrupt due to thermonuclear explosions during 
collapse (Fuller, Woosley, \& Weaver 1986). In fact, recent Newtonian 
simulations suggest that evolved zero-metallicity (Pop III) stars  
$\gtrsim 300 ~M_{\odot}$ do not disrupt but collapse with negligible mass loss 
(Fryer, Woosley, \& Heger 2001). This finding could be important since the 
first generation of stars may form in the range $10^2 - 10^3 ~M_{\odot}$ 
(Bromm, Coppi, \& Larson 1999; Abel, Bryan, \& Norman 2000). A combination of 
turbulent viscosity and magnetic fields likely will keep a spinning SMS in 
uniform rotation (Bisnovatyi-Kogan, Zel'dovich, \& Novikov 1967; Wagoner 1969; 
Zel'dovich \& Novikov 1971; Shapiro 2000; but see New \& Shapiro 2001 for an 
alternative). As they cool and contract, uniformly rotating SMSs reach the 
maximally rotating {\it mass-shedding limit}\ and subsequently evolve in a 
quasi-stationary manner along a mass-shedding sequence until reaching the 
instability point.  At mass-shedding, the matter at the equator moves in a 
circular geodesic with a velocity equal to the local Kepler velocity 
(Baumgarte \& Shapiro 1999). 

It is straightforward to understand the radial instability induced by general
relativity in a SMS by using an energy variational principle (Zel'dovich \& 
Novikov 1971; Shapiro \& Teukolsky 1983). Let $E=E(\rho_c)$ be the total 
energy of a momentarily static, spherical fluid configuration characterized by 
central mass density $\rho_c$.  The condition that $E(\rho_c)$ be an extremum 
for variations that keep the total rest mass and specific entropy distribution 
fixed is equivalent to the condition of hydrostatic equilibrium and 
establishes the relation between the equilibrium mass and central density:
\begin{equation} \label{1var}
\frac {\partial E}{\partial \rho_c} = 0 \ \Longrightarrow \ M_{\rm eq} 
= M_{\rm eq}(\rho_c) \\ ({\rm equilibrium}).
\end{equation}
The condition that the second variation of $E(\rho_c)$ be zero is the 
criterion for the onset of dynamical instability.  This criterion shows that 
the turning point on a curve of equilibrium mass vs. central density marks the 
transition from stability to instability:
\begin{equation} \label{2var}
\frac{{\partial}^2 E}{\partial {\rho_c}^2} = 0 \ \Longleftrightarrow \
\frac{\partial M_{\rm eq}}{\partial \rho_c} = 0 \\ 
({\rm onset\ of\ instability}).
\end{equation}

\begin{figure}[tb]
\includegraphics[angle=-90,width=11cm]{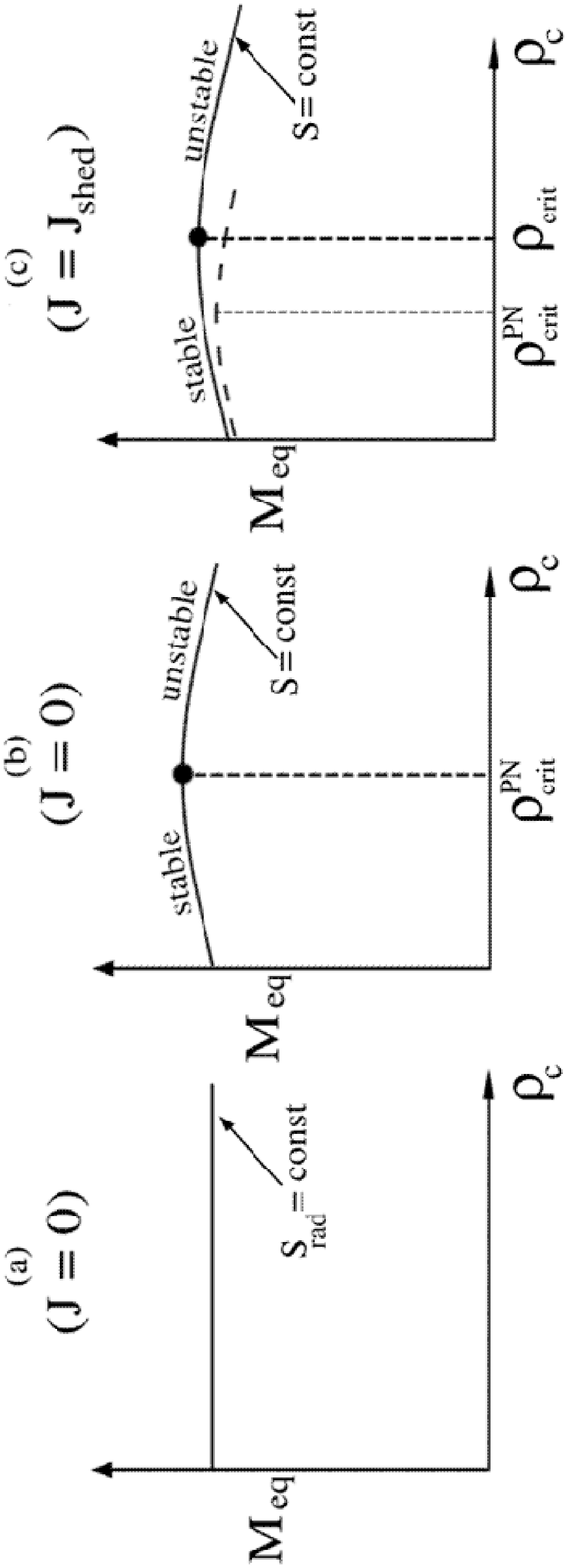}
\caption{A sketch of mass versus central density along an equilibrium
sequence of SMSs of fixed entropy. Panel ({\it a}) shows nonrotating, 
spherical Newtonian models supported by pure radiation pressure; ({\it b}) 
shows nonrotating, spherical PN models supported by radiation pressure plus 
thermal gas pressure; ({\it c}) shows rotating PPN models spinning at the 
mass-shedding limit.}
\label{sketch}
\end{figure}

Consider the simplest case of a spherical Newtonian SMS supported solely by 
radiation pressure and endowed with zero rotation.  This is an $n=3$, 
($\Gamma = 1+1/n= 4/3$) polytrope, with pressure 
\begin{equation} \label{eos1}
P = P_{\rm rad} = \frac{1}{3} a T^4 = K \rho^{\frac{4}{3}},
\end{equation}
where $K = K(s_{\rm rad})$ is a constant determined by the value of the 
(constant) specific entropy $s_{\rm rad} = \frac{4}{3} a T^3/n$  in the star. 
Here $T$ is the temperature, $n$ is the baryon number density, and $a$ is the 
radiation constant.  Consider a sequence of configurations with the same 
specific entropy but different values of central density. The total energy of 
each configuration is 
\begin{equation} \label{en1}
E(\rho_c) = U_{\rm rad} + W,
\end{equation}
where $U_{\rm rad}$ is the total internal radiation energy and $W$ is the 
gravitational potential energy.  Applying the equilibrium 
condition~(\ref{1var}) to this functional yields
$M_{\rm eq} = M_{\rm eq}(s_{\rm rad})$, i.e. the equilibrium mass depends only 
on the specific entropy and is independent of central density 
(see Fig.~\ref{sketch}{\it a}).  Applying the stability condition~(\ref{2var}) 
then shows that all equilibrium models along this sequence are marginally 
stable to collapse.

Now let us account for the effects of general relativity.  If we include the 
small (de-stabilizing) Post-Newtonian (PN) correction to the gravitational 
field, we must also include a comparable (stabilizing) correction to the 
equation of state arising from thermal gas pressure:
\begin{equation} \label{eos2}
P = P_{\rm rad} + P_{\rm gas} = \frac{1}{3} a T^4 + 2nkT,
\end{equation}
where we have taken the gas to be pure ionized hydrogen. Note that
$P_{\rm gas}/P_{\rm rad} = 8/(s_{\rm rad}/k) \ll 1$.
The energy functional of a star now becomes
\begin{equation} \label{en2}
E(\rho_c) = U_{\rm rad} + W + \Delta U_{\rm gas} + \Delta W_{\rm PN},
\end{equation}
where $\Delta U_{\rm gas}$ is the internal energy perturbation due to thermal
gas energy and $\Delta W_{\rm PN}$ is the PN perturbation to the gravitational 
potential energy.  Applying the equilibrium condition~(\ref{1var}) now yields
$M_{\rm eq} \approx M_{\rm eq}^{\rm Newt}$ times a slowly varying function of 
$\rho_c$  (see Fig.~\ref{sketch}{\it b}).  The turning point on the 
equilibrium curve marks the onset of radial instability; the marginally stable 
critical configuration is characterized by
\begin{eqnarray} \label{crit1}
\rho_{c,{\rm crit}} & = & 2 \times 10^{-3} M_6^{-7/2} 
{\rm gm\ cm^{-3}}, \nonumber \\
T_{c,{\rm crit}} & = & (3 \times 10^7) M_6^{-1}~{\rm K}, \\
(R/M)_{\rm crit} & = & 1.6 \times 10^3 M_6^{1/2}, \nonumber 
\end{eqnarray}
where $M_6$ denotes the mass in units of $10^6 ~M_{\odot}$.  (Here and 
throughout we adopt gravitational units and set $G=1=c$.)

Finally, let us consider a uniformly rotating SMS spinning at the 
mass-shedding limit (Baumgarte \& Shapiro 1999).  A centrally condensed object 
like an $n=3$ polytrope can only support a small amount of rotation before 
matter flys off at the equator. At the mass-shedding limit, the ratio of 
rotational kinetic to gravitational potential energy is only 
$T/|W| = 0.899 \times 10^{-2} \ll 1$. Most of the mass resides in a nearly 
spherical interior core, while the low-mass (Roche) envelope bulges
out in the equator: $R_{\rm eq}/R_{\rm pole} = 3/2$. When we include the
contribution of rotational kinetic energy to the energy functional, we must
now also include the effects of relativistic gravity to Post-Post-Newtonian 
(i.e. PPN) order, since both $T$ and $\Delta W_{\rm PN}$ scale with $\rho_c$
to the same power. The energy functional becomes
\begin{equation} \label{en3}
E(\rho_c) = U_{\rm rad} + W + \Delta U_{\rm gas} + \Delta W_{\rm PN}
+ \Delta W_{\rm PPN} + T
\end{equation}
Applying the equilibrium condition~(\ref{1var}), holding $M$, angular momentum 
$J$ and $s$ fixed, now yields $M_{\rm eq} \approx M_{\rm eq}^{\rm Newt}$ times 
a slowly varying function of $\rho_c$ (see Fig.~\ref{sketch}{\it c}).  If we 
restrict our attention to rapidly rotating stars with $M > 10^5 ~M_{\odot}$ 
the influence of thermal gas pressure is unimportant in determining the 
critical point of instability.  The turning point on the equilibrium curve 
then shifts to higher density and compaction than the critical values for 
nonrotating stars, reflecting the stabilizing role of rotation:
\begin{eqnarray} \label{crit2}
\rho_{c,{\rm crit}} & = & 0.9 \times 10^{-1} M_6^{-2} 
{\rm gm\ cm^{-3}}, \nonumber \\
T_{c,{\rm crit}} & = & (9 \times 10^7) M_6^{-1/2}~{\rm K}, \\
(R_{\rm pole}/M)_{\rm crit} & = & 427, \nonumber \\
(J/M^2)_{\rm crit} & = & 0.97. \nonumber 
\end{eqnarray}
The actual values quoted above for the critical configuration were determined 
by a careful numerical integration of the general relativistic equilibrium 
equations for rotating stars (Baumgarte \& Shapiro 1999); they are in close 
agreement with those determined analytically by the variational treatment. The 
numbers found for the nondimensional critical compaction and angular momentum 
are quite interesting.  First, they are universal ratios that are independent 
of the mass of the SMS. This means that a single relativistic simulation will 
suffice to track the collapse of a marginally unstable, maximally rotating SMS 
of arbitrary mass.  Second, the large value of the critical radius shows 
that a marginally unstable configuration is nearly Newtonian at the onset of 
collapse. Third, the fact that the angular momentum parameter of the critical 
configuration $J/M^2$ is below unity suggests that, in principle, the entire 
mass and angular momentum of the configuration could collapse to a rotating 
black hole without violating the Kerr limit for black hole spin (but see 
below!).

There are several plausible outcomes that one might envision {\it a priori}\ 
for the dynamical collapse of a uniformly rotating SMS once it reaches the 
marginally unstable critical point identified above. It could collapse to a 
clumpy, nearly axisymmetric disk, similar to the one arising in the Newtonian 
SPH simulation of Loeb \& Rasio (1994) for the isothermal ($\Gamma=1$) 
implosion of an initially homogeneous, uniformly rotating, low-entropy cloud. 
Alternatively, the disk might develop a large-scale,  nonaxisymmetric bar. 
After all, the onset of a dynamically unstable bar mode in a spinning 
equilibrium star occurs when the ratio $T/|W| \approx 0.27$ (see, e.g., 
Chandrasekhar 1969 and Lai, Rasio, \& Shapiro 1993 for Newtonian treatments
and Saijo et al. 2001 and Shibata, Baumgarte, \& Shapiro 2000a for simulations 
in general relativity). Since $T/|W|$ is $0.899 \times 10^{-2}$ at the onset 
of collapse and scales roughly as $R^{-1}$ during collapse due to conservation 
of mass and angular momentum, this ratio climbs above the dynamical bar 
instability threshold when the SMS collapses to $R/M \approx 20$, well before 
the horizon is reached.  The growth of a bar might begin at this point.  
Indeed, a weak bar forms in simulations of rotating supernova core collapse 
(Rampp, M\"uller, \& Ruffert 1998; Brown 2001), but here the equation of state 
stiffens ($\Gamma > 4/3$) at the end of the collapse, triggering a bounce and 
thereby allowing more time for the bar to develop.  A rapidly rotating 
unstable SMS might not form a disk at all, but instead collapse entirely to a 
Kerr black hole; not surprisingly, a nonrotating spherical SMS has been shown 
to collapse to a Schwarzschild black hole (Shapiro \& Teukolsky 1979). 
Alternatively, the unstable rotating SMS might collapse to a rotating black 
hole {\it and}\ an ambient disk.

\begin{figure}
\begin{tabular}{rl}
\includegraphics[width=5cm]{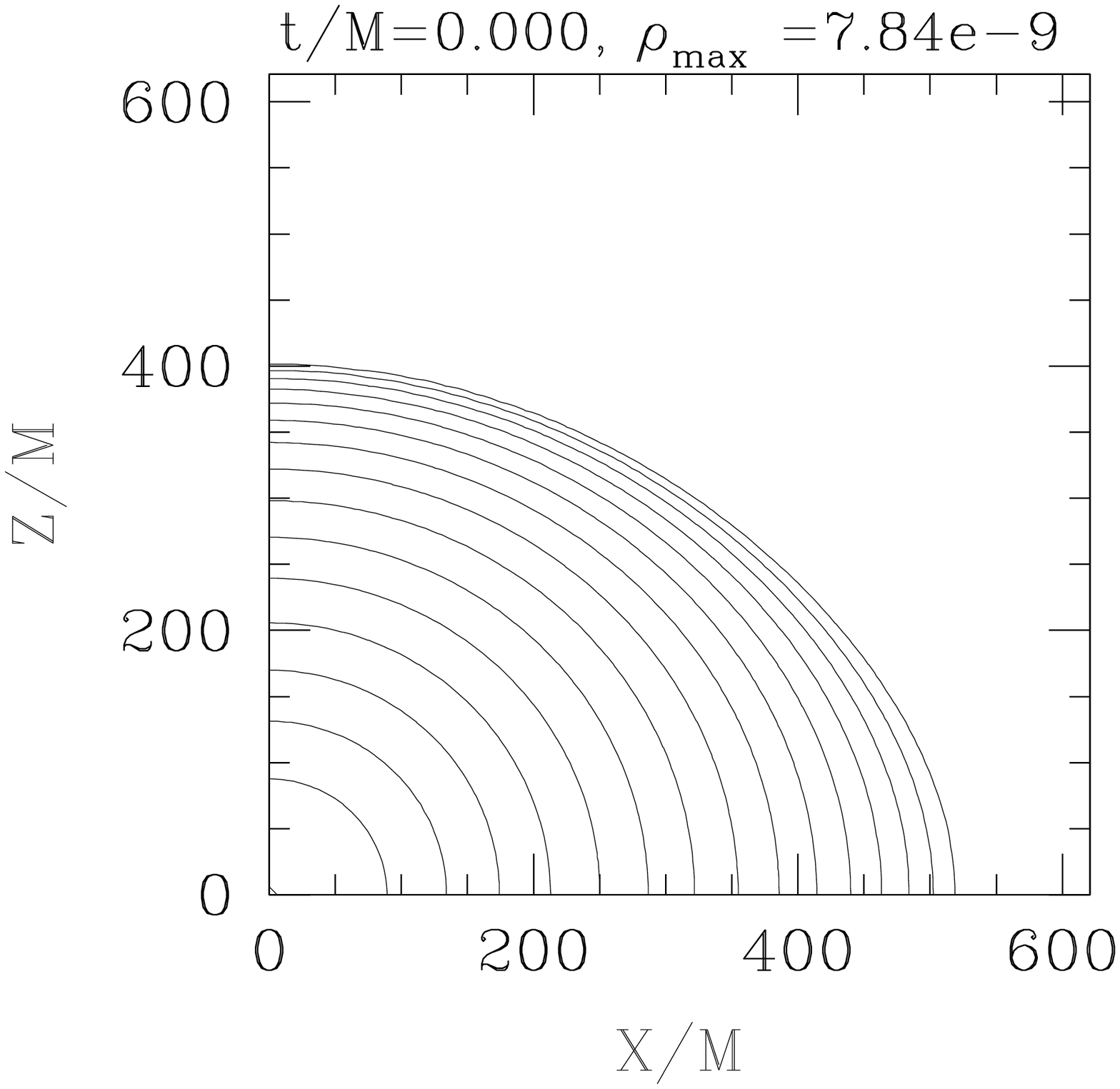}
&\includegraphics[width=5cm]{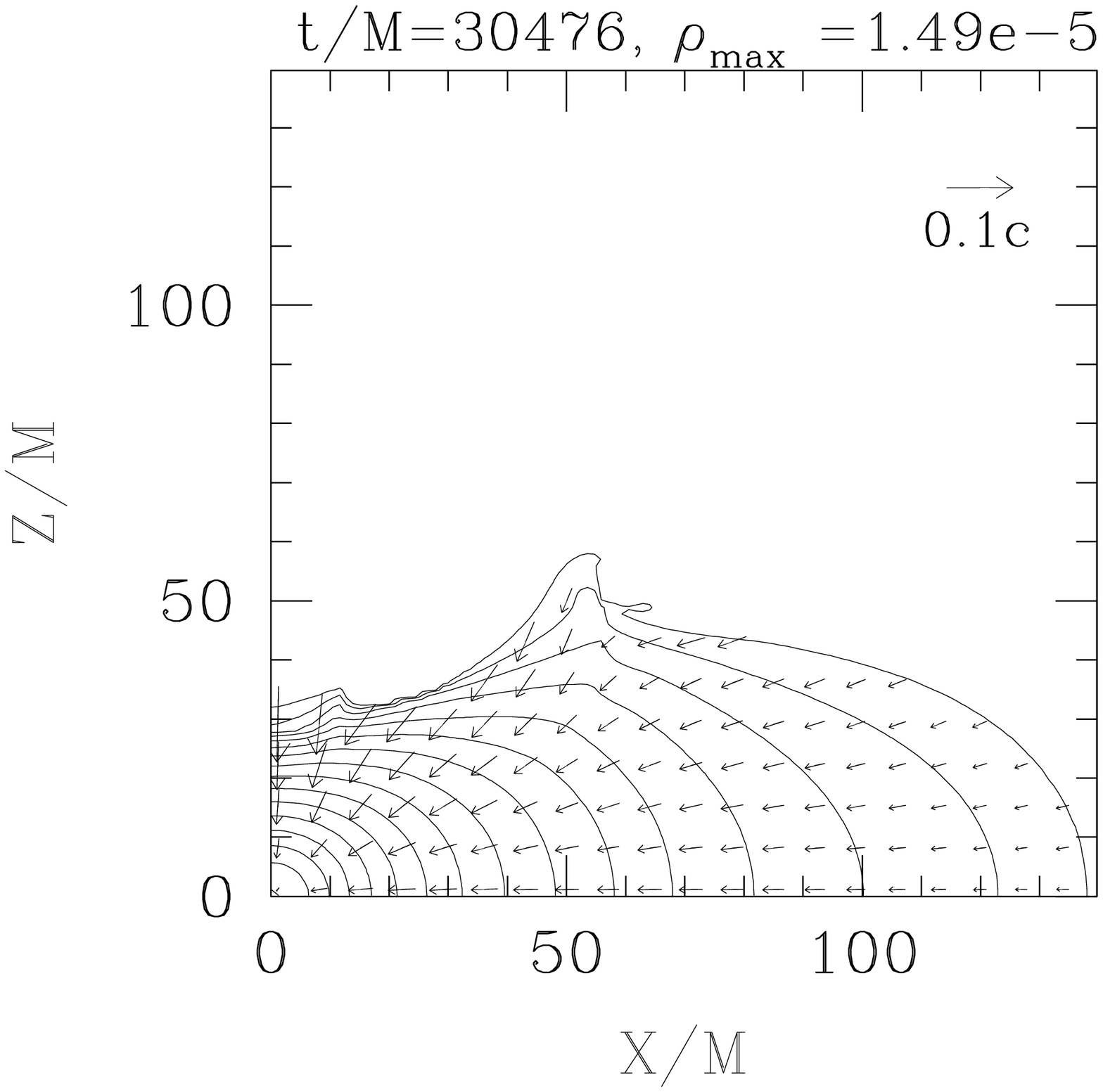}\\
\includegraphics[width=5cm]{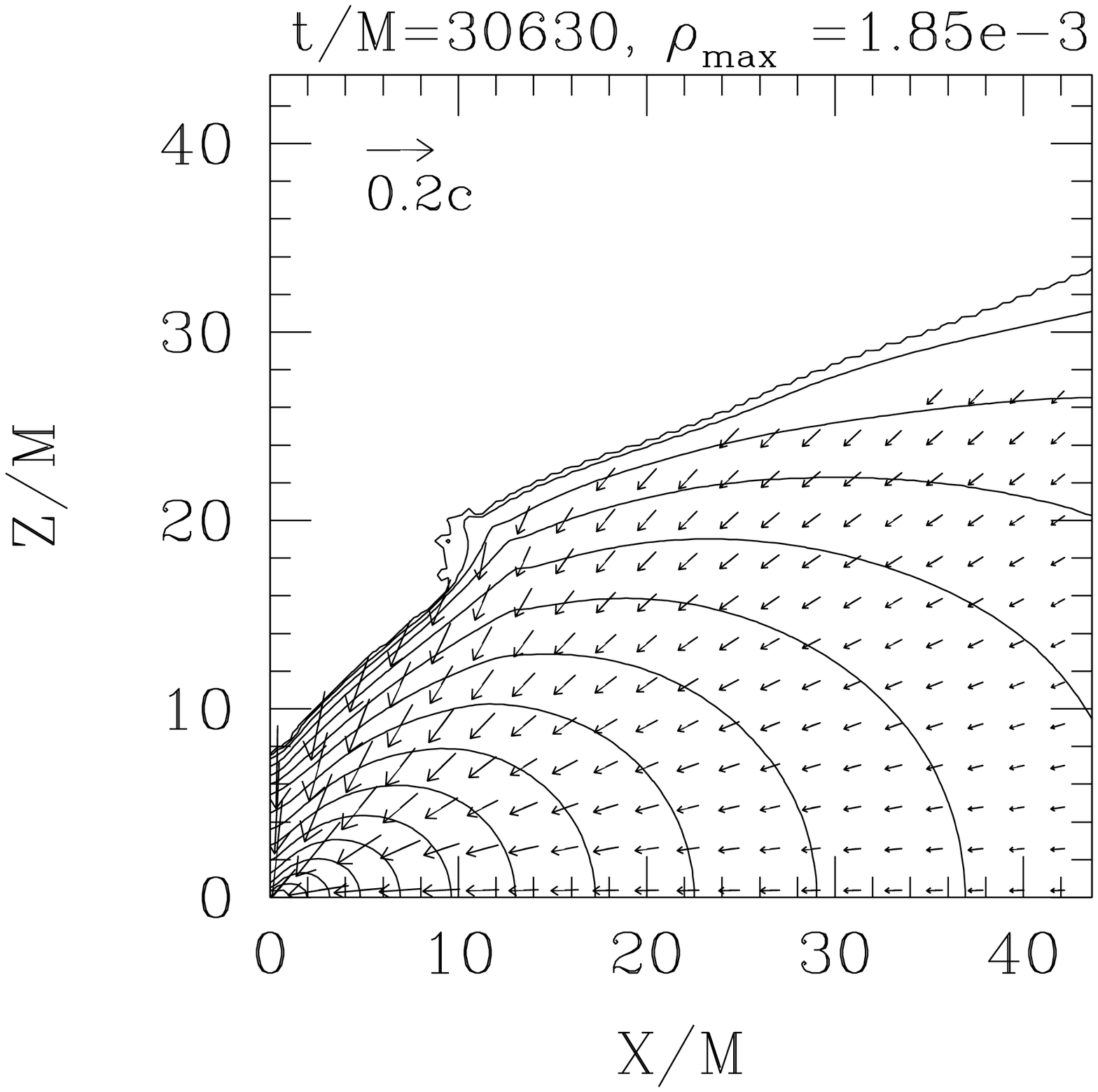}
&\includegraphics[width=5cm]{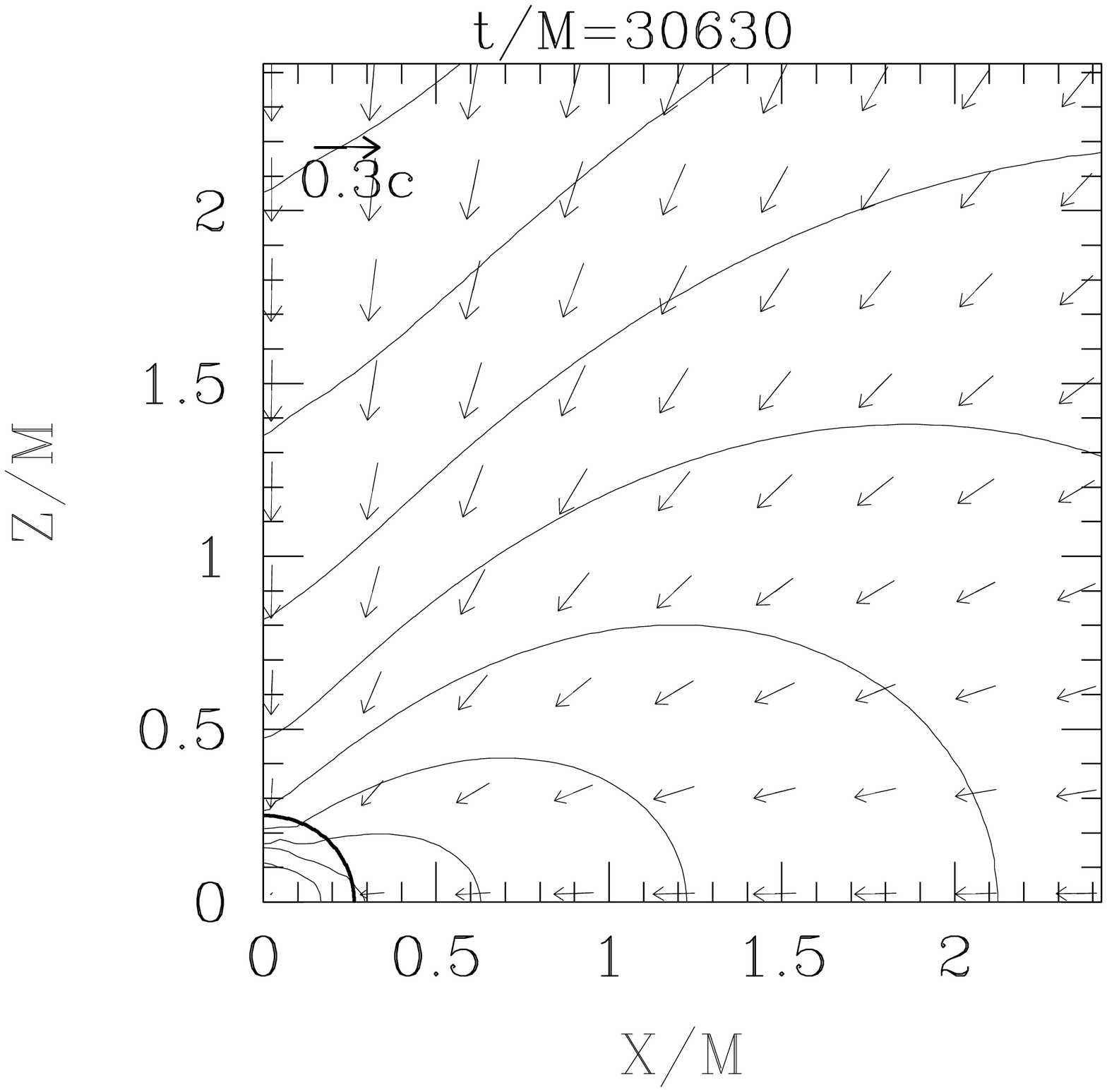}
\end{tabular}
\caption{Snapshots of density and velocity profiles during the implosion
of a marginally unstable SMS of arbitrary mass $M$ rotating uniformly at 
break-up speed at $t=0$.  The contours are drawn for $\rho/\rho_{\rm max}
=10^{-0.4j}~(j=0-15)$, where $\rho_{\rm max}$ denotes the maximum density at 
each time.  The fourth figure is the magnification of the third one in the 
central region: the thick solid curve at $r \approx 0.3M$ denotes the location 
of the apparent horizon of the emerging SMBH. (From Shibata \& Shapiro 2002.)}
\label{SMS}
\end{figure}

Two recent simulations have resolved the fate of a marginally unstable, 
maximally rotating SMS of arbitrary mass $M$. Saijo et al. (2002) followed the 
collapse in full $3D$, assuming PN theory. They tracked the implosion up to 
the point at which the central spacetime metric begins to deviate appreciably
from flat space at the stellar center. They found that the massive core 
collapses homologously during the Newtonian epoch of collapse, and that 
axisymmetry is preserved up to the termination of the integrations.  This 
calculation motivated Shibata \& Shapiro (2002) to follow the collapse in full 
general relativity by assuming axisymmetry from the beginning (see 
Fig.~\ref{SMS}).  They found that the final object is a Kerr-like black hole 
surrounded by a disk of orbiting gaseous debris.  The final black hole mass 
and spin were determined to be  $M_h/M \approx 0.9$ and $J_h/M_h^2 \approx 
0.75$. The remaining mass goes into the disk of mass $M_{\rm disk}/M \approx 
0.1$.  A disk forms even though the total spin of the progenitor star is 
safely below the Kerr limit.  This outcome results from the fact that the dense 
inner core collapses homologously to form a central black hole, while the 
diffuse outer envelope avoids capture because of its high angular momentum. 
Specifically, in the outermost shells, the angular momentum per unit mass $j$, 
which is strictly conserved on cylinders, exceeds $j_{\rm ISCO}$, the specific 
angular momentum at the innermost stable circular orbit about the final hole. 
This fact suggests how the final black hole and disk parameters can be 
calculated {\it analytically}\ from the initial SMS density and angular 
momentum distribution (Shapiro \& Shibata 2002). The result applies to the 
collapse of {\it any}\ marginally unstable $n=3$ polytrope at mass-shedding. 
Maximally rotating stars which are characterized by stiffer equations of state 
and smaller $n$ (higher $\Gamma$) do not form disks, typically, since they are
more compact and less centrally condensed at the onset of collapse (Shibata, 
Baumgarte, \& Shapiro 2000b).

The above calculations show that a SMBH formed from the collapse of a 
maximally rotating SMS is always born with a ``ready-made'' accretion disk. 
This disk might provide a convenient source of fuel to power the central 
engine.  The calculations also show that the SMBH will be born rapidly 
rotating.  This fact is intriguing in light of suggestions that observed SMBHs 
rotate rapidly (e.g., Wilms et al. 2001; Elvis, Risaliti, \& Zamorani 2002).

\section{Collapse of Collisionless Matter to a SMBH}

\begin{figure}[t]
\begin{tabular}{rl}
\includegraphics[angle=90,width=5.5cm]{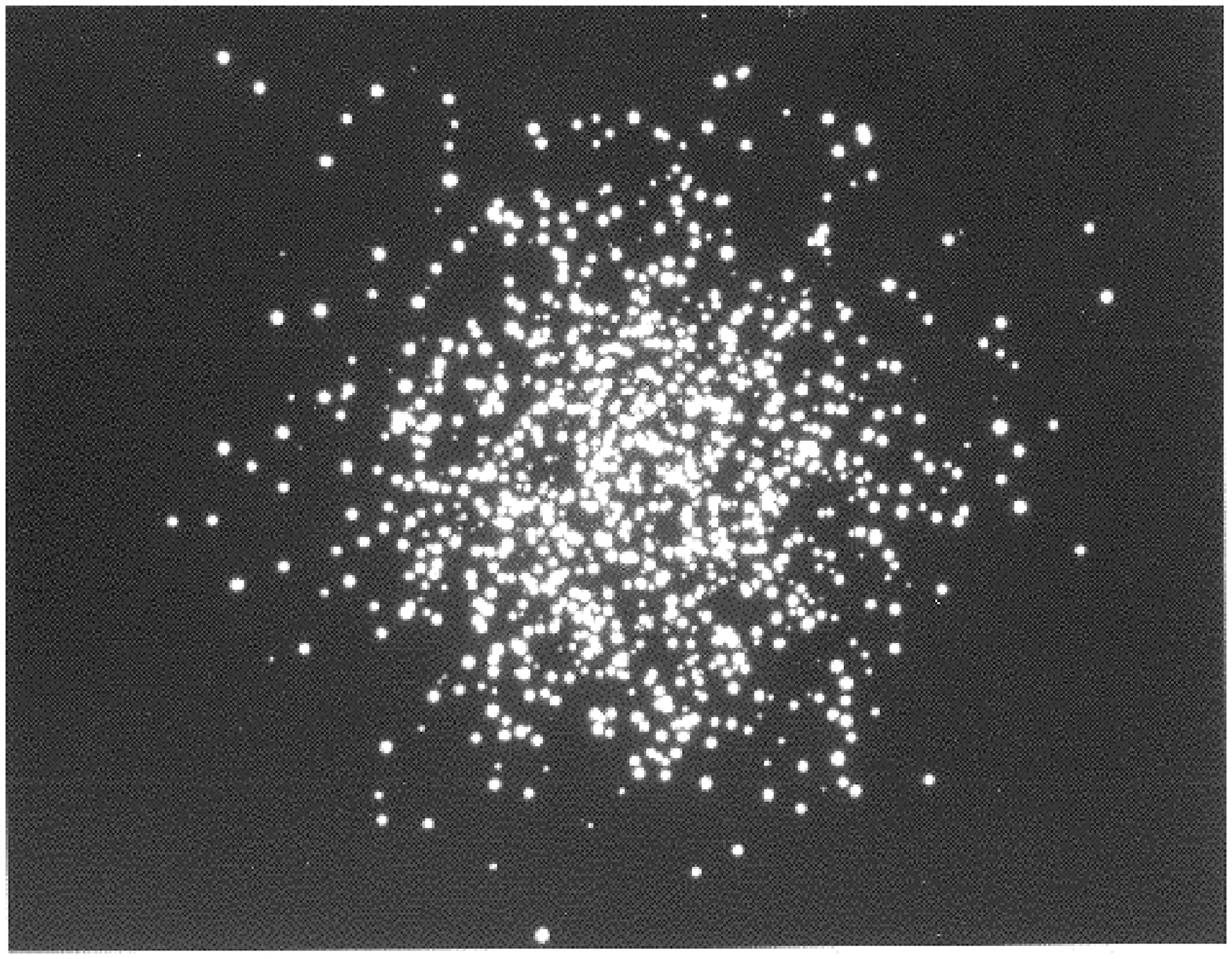}
&\includegraphics[angle=90,width=5.5cm]{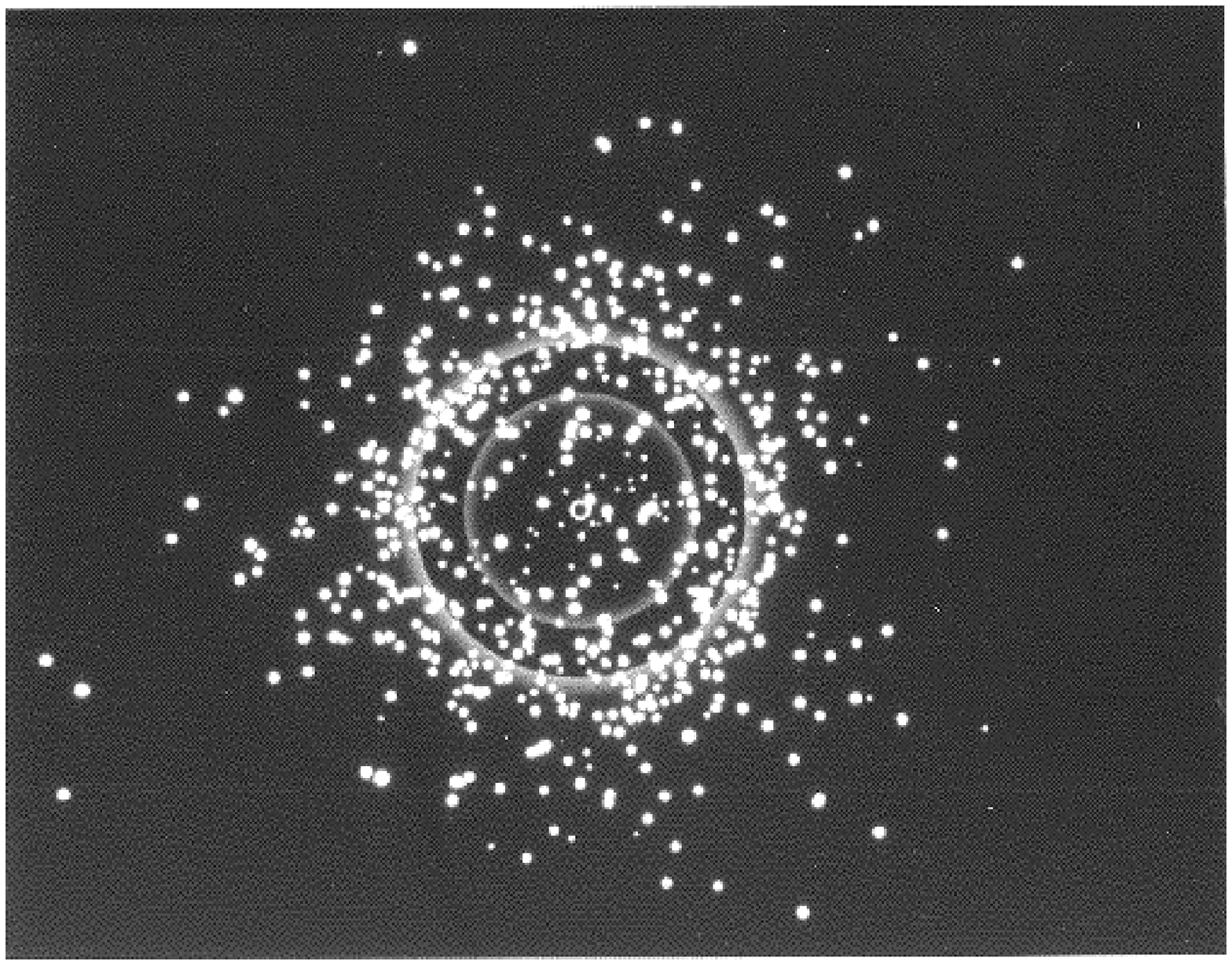}\\
\includegraphics[angle=90,width=5.5cm]{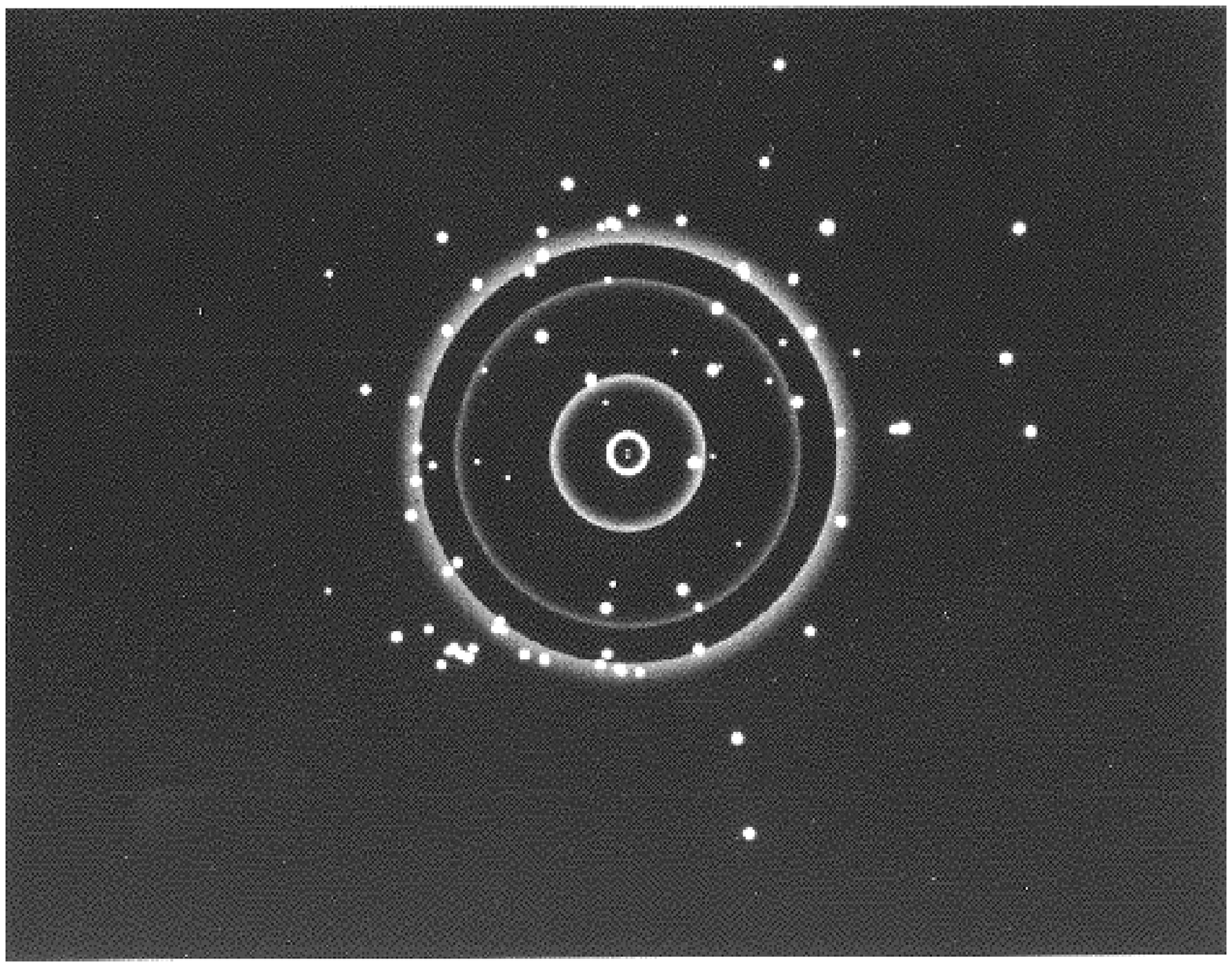}
&\includegraphics[angle=90,width=5.5cm]{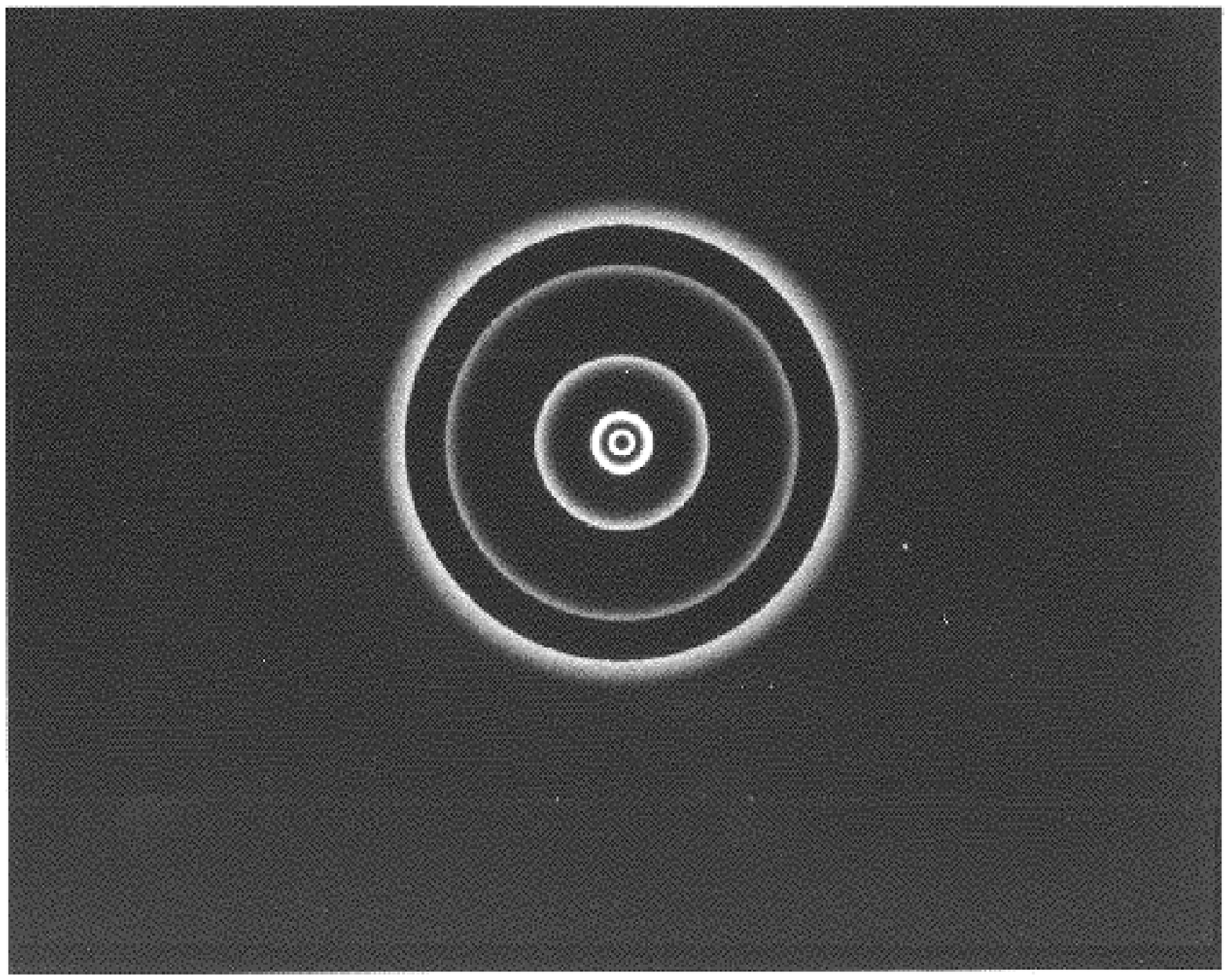}
\end{tabular}
\caption{The collapse of a marginally unstable gas of collisionless
particles of arbitrary mass $M$ 
which at $t=0$ obeys a Maxwell-Boltzmann distribution 
function with an areal radius $R/M = 9.0$. Spherical flashes of light 
are used to probe the spacetime geometry; at 
late times the light rays are trapped by the gravitational field. Their 
trajectories help locate the black hole event horizon, which in this 
example eventually reaches $r_s/M =2$  and encompasses all the 
matter. (After Shapiro \& Teukolsky 1988.)}
\label{max}
\end{figure}

Zel'dovich \& Podurets (1965) speculated that sufficiently compact, 
relativistic clusters of collisionless particles (e.g., relativistic star 
clusters) would be unstable to gravitational collapse. It has taken 
considerable theoretical effort to prove that this speculation is correct.
For a given distribution function $f = f(E)$, one  can construct a sequence of
spherical equilibrium clusters parametrized by the gravitational redshift 
at the cluster center, $z_c$. One can then plot a curve of fractional binding 
energy $E_{\rm b}/M_0 = 1 - M/M_0$. vs. $z_c$ along the sequence.  Linear 
perturbation theory, implemented via  a variational principle and trial 
functions, shows that the onset of radial instability occurs near the first 
turning point on such a binding energy curve (Ipser \& Thorne 1968; Ipser 
1969; Fackerell 1970). A rigorous theorem has been proven that spherical 
equilibrium configurations are stable, at least up to the first turning point 
on the binding energy curve (Ipser 1980). Numerical simulations have shown that
all spherical configurations beyond the first turning point are dynamically
unstable (Shapiro \& Teukolsky 1985a,b,c; 1986).  Most significantly, these 
simulations have tracked the nonlinear evolution of unstable spherical and 
axisymmetric clusters, including those with rotation (Abrahams et al. 1994; 
Shapiro, Teukolsky, \& Winicour 1996), and have determined their final fate
(see Fig.~\ref{max}). This computational enterprise (``relativistic stellar 
dynamics on a computer;'' see Shapiro \& Teukolsky 1992 for a review and 
references) has demonstrated the following: 
\begin{enumerate}\normalsize
\item the dynamical collapse of a collisionless cluster leads to the formation
of a Kerr black hole whenever $J/M^2 \lesssim 1$;
\item collapse leads instead to a bounce followed by virialization 
of the cluster by relativistic violent relaxation whenever $J/M^2 \gtrsim 1$;
\item in extreme core-halo systems, collapse leads to a new stationary,
equilibrium system consisting of a central black hole surrounded by an
extensive, nearly Newtonian halo containing most of the mass.
\end{enumerate}\normalsize

\begin{figure}[t]
\begin{tabular}{rl}
\includegraphics[height=6cm]{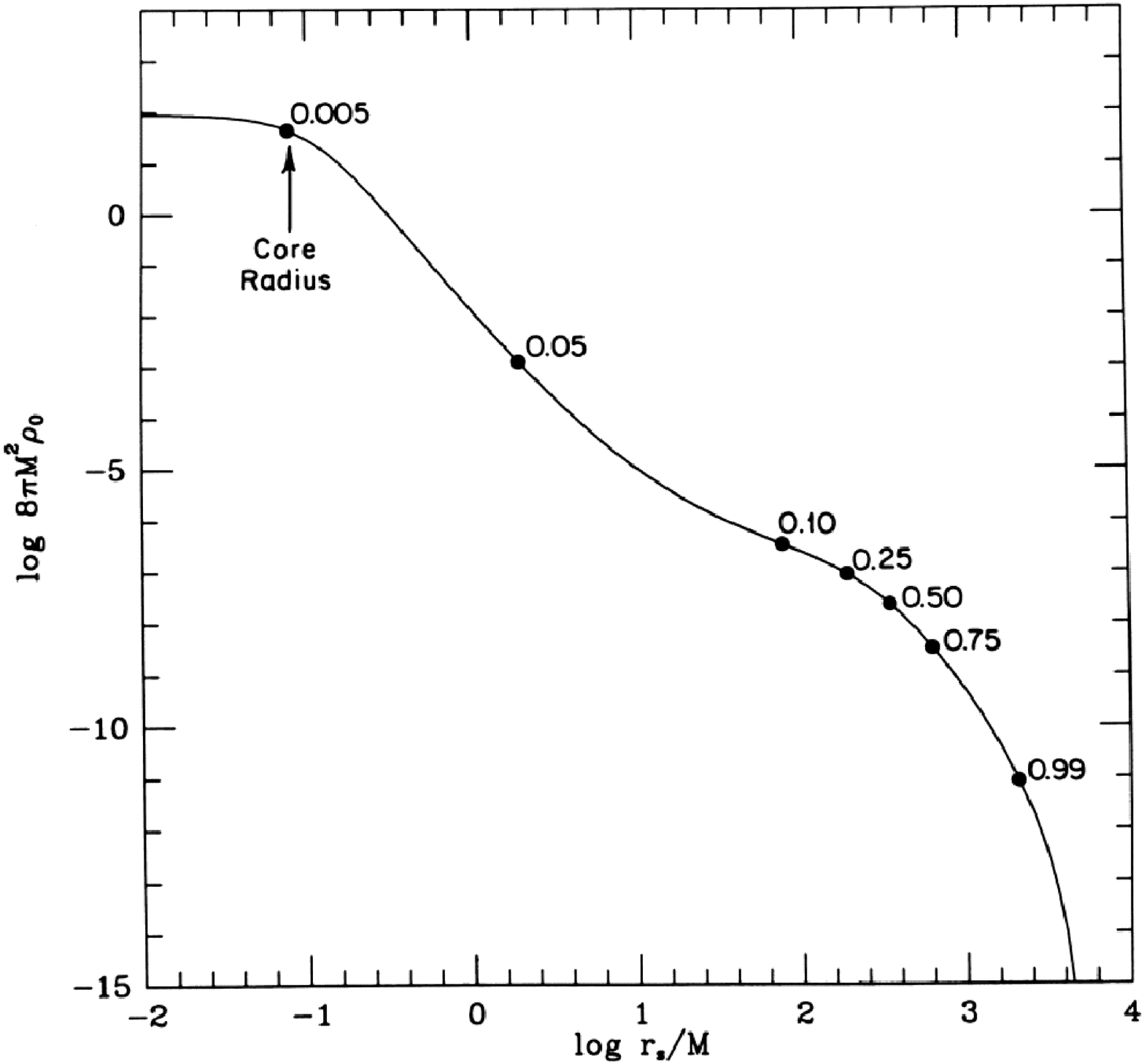}
&\includegraphics[height=6cm]{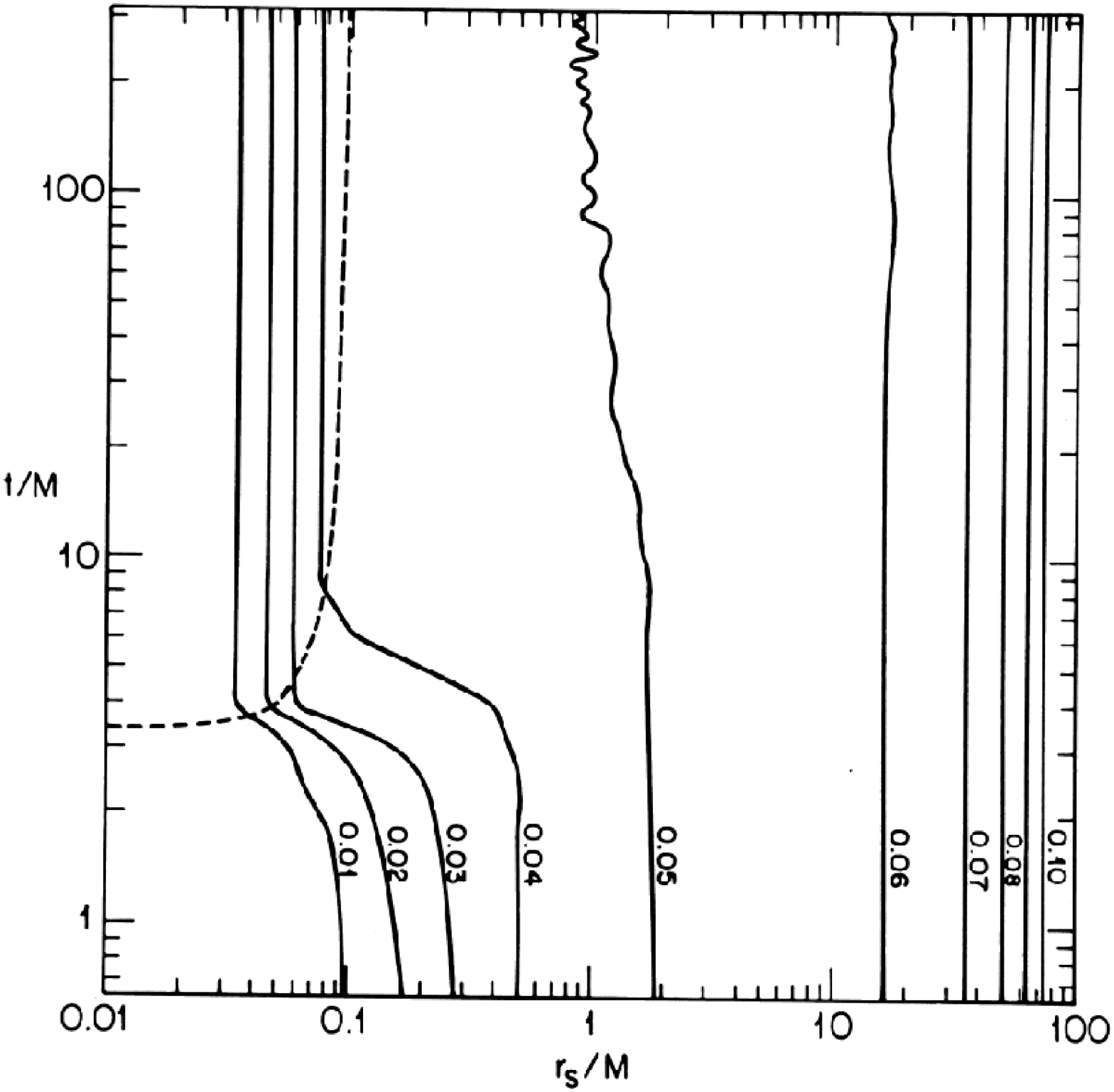}
\end{tabular}
\caption{An extreme core-halo configuration of arbitrary mass $M$ constructed 
from an $n=4$ relativistic polytropic distribution function. The {\it left}\ 
panel shows the initial equilibrium rest-mass density profile.  The points 
label the interior rest-mass fraction. This cluster has a highly relativistic 
core and an extensive Newtonian halo and is marginally unstable to collapse.  
The {\it right}\ panel is a spacetime diagram showing the worldlines of 
imaginary Lagrangian matter tracers. The dashed line shows the event horizon 
of the black hole that forms at the center. Note that the central core and its 
surroundings undergo collapse but that $95 \%$ of the cluster mass settles 
into stable dynamical equilibrium about the central hole. (From Shapiro \& 
Teukolsky 1986.)}
\label{prof}
\end{figure}

\begin{figure}
\centering
\includegraphics[width=10cm]{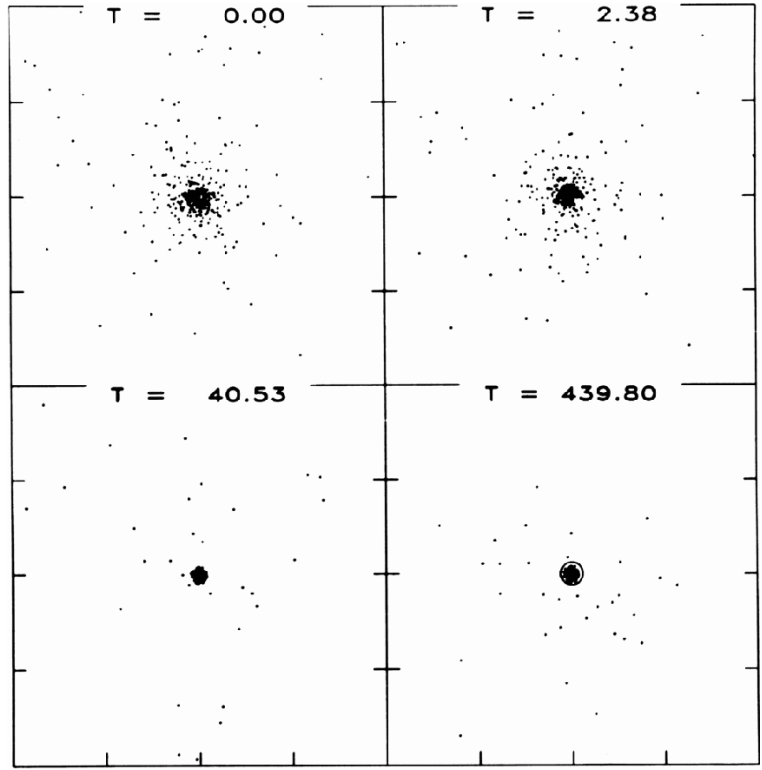}
\caption{Snapshots of the central particle distribution inside $r_s/M = 2$ at 
selected times during the collapse described in Figure~\ref{prof}. The cluster 
does not evolve appreciably after $t/M = 40$. The circle in the last frame 
shows the black hole event horizon at $r_s/M = 0.1$.  (From Shapiro \& 
Teukolsky 1986.)}
\label{quas}
\end{figure}

Conclusion (3) is very tantalizing as a plausible route for forming SMBHs (see 
Fig.~\ref{prof} and \ref{quas}).  But a key question remains: under what 
circumstances can a cluster be driven to a dynamically unstable, relativistic 
state to trigger such a collapse?

\begin{figure}
\centering
\includegraphics[width=7cm]{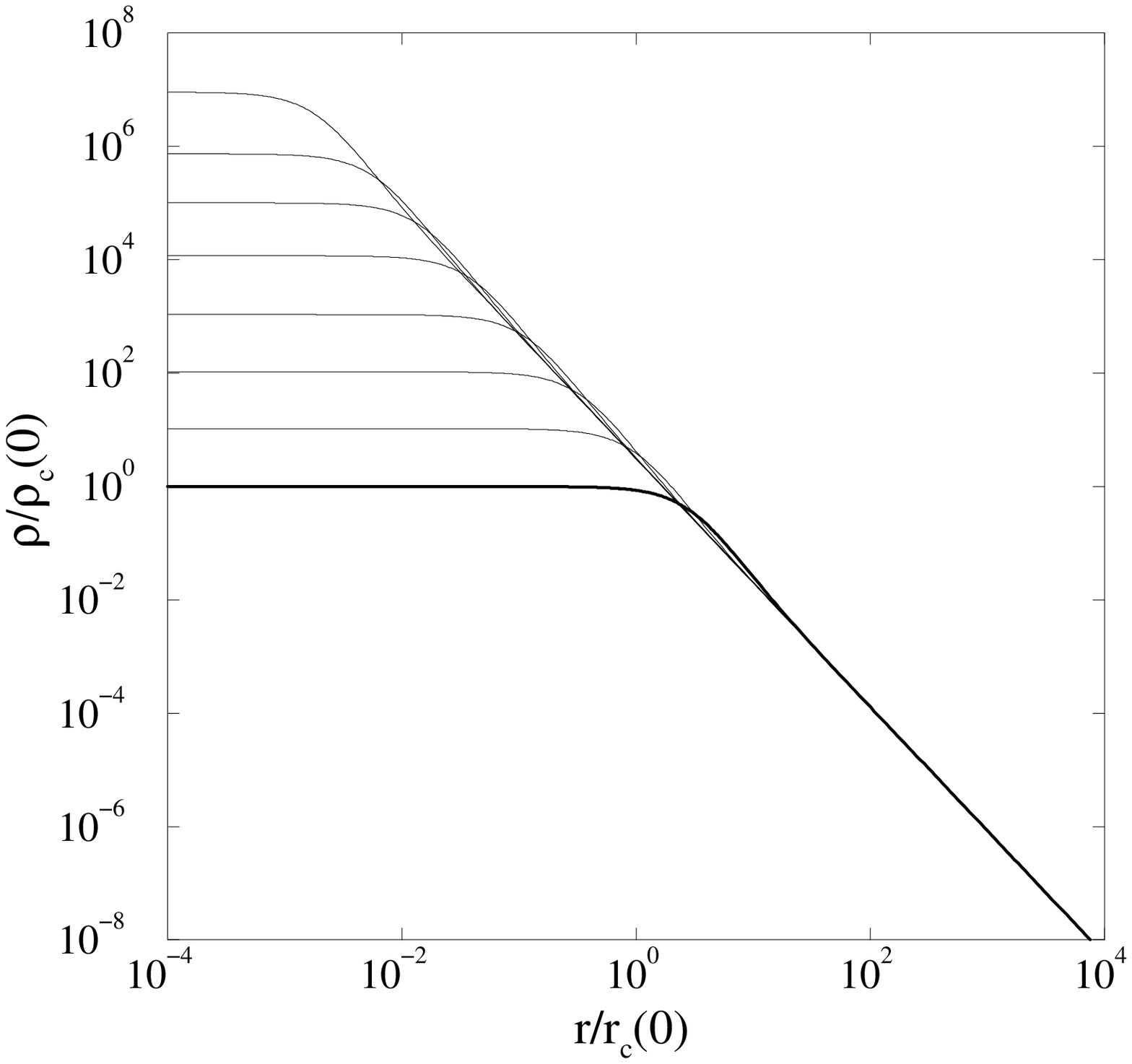}
\caption{Snapshots of the self-similar density profile at selected times
during {\it secular}\ core collapse of a nearly collisionless Newtonian 
cluster (the gravothermal catastrophe). The thick line shows the profile 
at $t=0$ and successive profiles with higher central densities 
correspond to later times. (From Balberg, Shapiro, \& Inagaki 2002.)}
\label{gravo}
\end{figure}

One possibility may involve the ``gravothermal catastrophe,'' the runaway core 
contraction on a relaxation timescale of a stable, virialized cluster due to 
the perturbative influence of collisions (Chandrasekhar 1942; Lynden-Bell \& 
Wood 1968; see Spitzer 1975, 1987 and Lightman \& Shapiro 1978 for reviews and 
references). Collisional scattering is a source of kinetic energy transport 
(heat conduction).  Self-gravitating, nearly collisionless clusters with 
$t_d/t_r \ll 1$ have negative heat capacity: as their high-temperature cores 
lose energy to their low-temperature halos by heat conduction, the cores 
contract and, in accord with the virial theorem, become hotter still, leading 
to a thermal runaway.  The result is that the clusters undergo homologous core 
contraction (Lynden-Bell \& Eggleton 1980), as depicted in Figure~\ref{gravo}
for Newtonian clusters composed of identical particles.  Contraction to a 
singular state is complete in a time $\approx 300 t_r(0)$, where the central 
relaxation timescale $t_r(0)$ is measured from an arbitrary initial time $t=0$ 
(Cohn 1979).  Specifically, as $t \rightarrow 300 t_r(0)$, the central  
density as well as the  the central redshift, potential and velocity 
dispersion (temperature) all blow up: $\rho_c \rightarrow \infty$ and 
$z_c \sim \Phi_c \sim \upsilon_c^2 \rightarrow \infty$. At the same time the core 
radius and mass both shrink: $r_c \rightarrow 0$ and $M_c \rightarrow 0$. The 
required number of relaxation timescales to reach this singular state, as well 
as the $r^{-2.2}$ fall-off in the halo density profile, are nearly independent 
of the velocity dependence of the collision cross section (Balberg \& Shapiro, 
unpublished). This gravothermal catastrophe can, in principle, drive a core to 
a highly relativistic state, at which point it could become
dynamically unstable to catastrophic collapse on a dynamical timescale.

Detailed numerical calculations have been performed for several scenarios to 
test whether the gravothermal catastrophe can trigger the formation of SMBHs. 
We will summarize three of them below.

\subsection{Gravothermal evolution of a cluster of {\it compact}\ stars}

Here we describe simulations involving the evolution of clusters of compact 
stars (neutron stars or stellar-mass black holes) and the build-up of massive
black holes. This route has been discussed by several authors (e.g., 
Zel'dovich \& Podurets 1965; Rees 1984; Shapiro \& Teukolsky 1985c; Quinlan \& 
Shapiro 1987, 1989). We briefly summarize below the key results of the 
multi-mass Fokker-Planck calculations of Quinlan \& Shapiro (1989).

Consider a dense cluster of compact stars composed initially of identical 
neutron stars or black holes of mass $m_{\ast} = 1.4 ~M_{\odot}$ in virial 
equilibrium.  Take the central mass density to be 
$\rho_c \approx 10^8 ~M_{\odot} {\rm ~pc^{-3}}$ and the central velocity 
dispersion to be $\upsilon_c \approx 500 {\rm ~km ~s^{-1}}$.  Here gravothermal 
evolution is driven by the cumulative effect of repeated, distant, 
small-angle, gravitational (Coulomb) scatterings between the stars.  Binary 
formation is significant and is dominated by dissipative, 2-body capture by 
gravitational radiation. 

The simulations reveal that mass segregation causes significant departures 
from single-component homological evolution models.  For example, $\upsilon_c$ does 
not increase at the center and the cluster is not driven to a relativistic 
state. However, there is an inevitable build-up of massive BHs via successive 
binary mergers. The evolution is followed up to the formation of BHs of mass 
$M_{\rm BH} \gtrsim 100 ~M_{\odot}$, at which point the number of stars in the 
core become sufficiently small that the Fokker-Planck treatment breaks down. 
The intermediate mass BH binaries produced in this scenario would be prime 
sources of gravitational waves for the ground-based network of laser 
interferometers now under construction (LIGO, VIRGO, GEO, and TAMA) and
for the space-based interferometer currently being designed (LISA).

\subsection{Gravothermal evolution of a dense cluster of {\it ordinary}\ stars}

Here we discuss simulations that start from more typical initial conditions. 
These calculations treat the  gravothermal evolution of a dense cluster of 
ordinary, main-sequence stars (Spitzer \& Saslaw 1966; Colgate 1967; Sanders 
1970; Begelman \& Rees 1978; Lee 1987; Quinlan \& Shapiro 1990; Gao et al. 
1991; Portegies Zwart \& McMillan 2002). Here  we briefly summarize the  
multi-mass Fokker-Planck calculations of Quinlan \& Shapiro (1990), which are 
among the most detailed for galactic nuclei that do not assume the presence 
of a SMBH {\it a priori}.

Consider a dense galactic nucleus initially consisting of main-sequence stars 
with mass $m_{\ast} = 1.0 ~M_{\odot}$ in virial equilibrium.  Take the central 
mass density to be in the range $\rho_c \approx 10^6 - 10^8 ~M_{\odot} 
{\rm ~pc^{-3}}$ and the central velocity dispersion to be in the range 
$\upsilon_c \approx 100 - 400 {\rm ~km ~s^{-1}}$.  This velocity is below the escape 
velocity from the surface of the stars, so that collisions will lead to 
mergers and not disruptions. Collisions and mergers, stellar evolution and the 
formation of new stars from gas liberated by supernovae are included 
in the calculation, as are the formation of binaries by 3-body encounters and
the interaction between hard binaries and single stars. All of this activity
takes place in the context of the gravothermal evolution of the cluster,
which again is driven by the cumulative effect of repeated, distant,
small-angle, gravitational scattering of the stars. 

The outcome of the evolution is that stars with 
$m_{\ast} \gtrsim 100 ~M_{\odot}$ form easily, then merge and collapse to form
seed BHs with masses $M_{\rm BH} ~\approx 100 - 1000 ~M_{\odot}$ in a time
$t \lesssim 10^{10} {\rm ~yrs}$. The end result is the formation of a dense 
cluster of compact remnants comprised of intermediate mass black holes. The 
cluster is characterized by frequent binary mergers ($\gg 1$ per year when 
integrated throughout the visible universe of $\sim 10^{10}$ galaxies). These 
intermediate mass black hole binaries are again promising  sources of 
gravitational waves for the ground-based laser interferometers like LIGO and 
for the proposed space-based interferometer LISA.

\subsection{Gravothermal contraction of an SIDM halo to a relativistic state} 

Dark matter comprises about $ 90 \%$ of the matter in the universe.  The 
simplest description of dark matter which accounts for many  features of the 
large-scale structure of the universe is the ``cold dark matter'' (CDM) model, 
in which the dark matter particles are essentially collisionless. However,
the possibility that dark matter particles interact with each other strongly
and have a substantial scattering cross section has been revived recently 
(Spergel \& Steinhardt 2000) to explain several observations of dark matter
structures on the order of $\lesssim 1~{\rm Mpc}$. Dynamical studies confirm 
that halos formed from such ``self-interacting'' dark matter (SIDM) have 
flatter density cores in better agreement with the observations than the more 
cuspy profiles predicted by standard CDM.

\begin{figure}
\centering
\includegraphics[width=7cm]{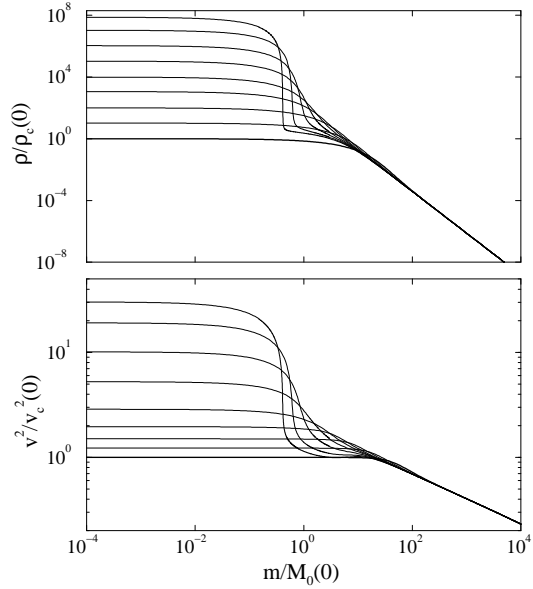}
\caption{Snapshots of the ({\it a}) density and ({\it b}) velocity dispersion 
profiles of an SIDM halo at selected times during gravothermal evolution.  The 
thick line shows the profile at $t=0$ and successive profiles with higher 
central densities correspond to later times. Bifurcation into a fluid inner 
core and a collisionless outer core is evident at late times.  (From Balberg 
et al. 2002.)}
\label{SIDM}
\end{figure}

Balberg \& Shapiro (2002) have recently demonstrated that SMBH formation may 
be an inevitable consequence of dynamical core collapse following the 
gravothermal catastrophe in SIDM halos. This conclusion follows from their 
earlier dynamical study (Balberg et al. 2002) which tracked the gravothermal 
evolution of a virialized, spherical SIDM halo by employing the fluid 
formalism of Lynden-Bell \& Eggleton (1980).  In the early universe, halos 
form with a Press-Schechter (1974) distribution. Typical halos are 
characterized by a central mass density 
$\rho_c \gtrsim  10^{-2} ~M_{\odot} {\rm ~pc^{-3}}$, a central velocity 
dispersion $\upsilon_c \gtrsim 100 {\rm ~km ~s^{-1}}$ and an elastic scattering
cross section $\sigma \gtrsim 0.1 {\rm ~cm^2 ~gm^{-1}}$. The ratio of the 
scattering mean free path $\lambda$ to the gravitational scale height $H$ 
everywhere satisfies the long mean free path inequality $\lambda / H \gg 1$ 
initially.  This results in homologous gravothermal contraction at first (see 
Fig.~\ref{gravo}).  However, in SIDM halos the interactions are large-angle 
scatterings between close neighbors, not cumulative, small-angle Coulomb 
scatterings by distant particles, as in star clusters.  Consequently, in 
contrast to star clusters, the inner core of an SIDM halo evolves into the 
short mean free path (fluid) regime where $\lambda / H \ll 1$. There is a
bifurcation of the halo into a fluid inner core surrounded by a collisionless
outer core and halo (see Fig.~\ref{SIDM}).  Continued heat conduction out of 
the inner core drives it to a relativistic state, which eventually becomes
unstable to collapse to a black hole.

The initial mass of the black hole will be $10^{-8}-10^{-6}$ of the total mass 
of the halo. Very massive SIDM halos form SMBHs with masses 
$M_{\rm BH} \gtrsim 10^6 ~M_{\odot}$ directly. Smaller halos believed to form
by redshift $z \approx 5$ produce seed black holes of mass $M_{\rm BH} \approx
10^2 - 10^3 ~M_{\odot}$ which can merge and/or accrete to reach the observed 
SMBH range. Significantly, this scenario for SMBH formation requires no 
baryons, no prior star formation and no other black hole seed mechanisms (cf. 
Ostriker 2000; Hennawi \& Ostriker 2002).

\section{Conclusions and Final Thoughts}

Relativistic gravitation --- general relativity --- induces a dynamical, 
radial instability to collapse in {\it all}\ forms of self-gravitating matter 
whenever the matter becomes sufficiently compact.  Plausible scenarios have 
been proposed that can trigger this instability to form SMBHs or their seeds.
SMBHs are believed to reside at the centers of quasars, AGNs, and many, if 
not all, normal galaxies with bulges, including the Milky Way.  According to 
recent observations, even globular clusters may contain black holes of 
intermediate mass $\sim 10^3 - 10^{4} ~M_{\odot}$ (Gebhardt, Rich, \& Ho 2002; 
Gerssen et al. 2002), although alternative interpretations have been proposed 
(Baumgardt et al. 2003).  Explaining the origin of SMBHs is thus a 
fundamental, unresolved issue of modern cosmology and structure formation.  
Some of the scenarios proposed for forming SMBHs are ``hydrodynamical'' in 
nature, as in the collapse of fluid SMSs to SMBHs. Some of them are ``stellar 
dynamical,'' as in the collapse of a relativistic cluster of collisionless 
particles or compact stars. Still others are ``hybrids,'' as in the case of 
the collapse of massive stars, built-up by collisions and binary mergers in 
dense clusters undergoing gravothermal contraction; or the case of the 
collapse of the fluid inner core of an SIDM halo, following its gravothermal 
contraction to a relativistic state. The challenge in exploring the competing 
scenarios is that the different physical regimes characterizing them  are 
described by very different sets of equations, requiring very different 
numerical techniques for solution.  Yet numerical simulations have begun to 
explore many of the proposed routes.

At the present time we do not know for certain what is the dominant route by 
which observed SMBHs are formed: Are they born supermassive, or do they grow 
supermassive from small seeds?  Do the seed black holes grow by merger, by gas 
accretion or both?  Do the first black holes arise from the collapse of 
ordinary baryonic matter, collisionless dark matter, or from some more exotic 
form of mass-energy (e.g., scalar fields?  gravitational waves?).

The growth of black hole seeds by gas accretion is supported by the 
consistency between the total energy density in QSO light and the BH mass 
density in local galaxies, adopting a reasonable accretion
rest-mass--to--energy conversion efficiency (Soltan 1982; Yu \& Tremaine 
2002).  But quasars have been discovered out to redshift $z \approx 6$, so it 
follows that the first SMBHs must have formed by $z_{\rm BH} \gtrsim 6$ or 
within $t_{\rm BH} \lesssim 10^9$ yr after the Big Bang. This timescale 
provides a tight constraint on SMBH formation.  For example, if SMBHs indeed 
grew by accretion, black hole seeds of mass $ \gtrsim 10^5 ~M_{\odot}$ must 
have formed by $ z \approx 9$ to have had sufficient time to reach a mass of 
$\sim 10^9 ~M_{\odot}$ by $z \approx 6$ (Gnedin 2001). 

The correlations $M_{\rm BH} \propto L_{\rm bulge}$ (Kormendy \& Richstone 
1995) and $M_{\rm BH} \propto \upsilon_c^4$ (Gebhardt et al. 2000; Ferrarese \& 
Merritt 2000) inferred for galaxies provide important additional constraints. 
For example, SMBH formation by mergers of smaller seed holes during the 
hierarchical build-up of galaxies can account for these scaling laws (Haehnelt 
\& Kauffmann 2000).  But some observations suggest that SMBHs spin rapidly.
This conclusion might restrict the significance of merger scenarios, since 
black holes are typically spun down by repeated mergers (Hughes \& Blandford 
2003). On the other hand, a single final merger with a binary companion of 
comparable mass could drive the spin of a black hole back up to a large value. 

Further observations, including the detection of gravitational waves from 
distant coalescing black hole binaries, might establish the evolutionary 
tracts and merging histories of SMBHs and help identify their principle 
formation mechanism.

This work was supported in part by NSF Grants PHY-0090310 and PHY-0205155 and 
NASA Grants NAG5-8418 and NAG5-10781 at the University of Illinois at 
Urbana-Champaign.

\end{document}